%% file: iclr2026_conference.tex
\documentclass{article} 
\usepackage{iclr2026_conference,times}

\input{math_commands.tex}

\usepackage{graphicx}
\usepackage{url}
\usepackage{xcolor}
\usepackage{algorithm}
\usepackage{algpseudocode}
\usepackage{amsmath}
\usepackage{booktabs}
\usepackage{pifont}
\usepackage{subcaption}
\usepackage{multirow}
\usepackage{listings}
\usepackage{enumitem}

\definecolor{softred}{HTML}{FFD9D9}
\definecolor{softgreen}{HTML}{C3F7C8}

\title{CR-Bench: Evaluating the Real-World Utility of AI Code Review Agents}

\author{Kristen Pereira\thanks{Equal contribution} , Neelabh Sinha\footnotemark[1] , Rajat Ghosh, Debojyoti Dutta \\
Nutanix, Inc. \\
\texttt{\{kristen.pereira, neelabh.sinha,} \\ 
\texttt{rajat.ghosh, debojyoti.dutta\}@nutanix.com}
}

\iclrfinalcopy 
\begin{document}

\maketitle

\begin{abstract}

Recent advances in frontier large language models have enabled code review agents that operate in open-ended, reasoning-intensive settings. However, the lack of standardized benchmarks and granular evaluation protocols makes it difficult to assess behavior of code review agents beyond coarse success metrics, particularly for tasks where false positives are costly. To address this gap, we introduce \texttt{CR-Bench}, a benchmarking dataset, and \texttt{CR-Evaluator}, a fine-grained evaluation pipeline for code review agents. Using these tools, we conduct a preliminary study evaluating both a single-shot agent and a Reflexion-based agent across two frontier models. We find that code review agents can exhibit a low signal-to-noise ratio when designed to identify all hidden issues, obscuring true progress and developer productivity when measured solely by resolution rates. Our analysis identifies the hidden trade-off between issue resolution and spurious findings, revealing a frontier that constrains effective agent design. Together, \texttt{CR-Bench} and \texttt{CR-Evaluator} provide a timely foundation for studying and developing code review agents as LLM-based systems transition from controlled benchmarks to real-world software engineering workflows.


 
\end{abstract}

\input{content/introduction}

\input{content/related_work}
\input{content/problem_definition}
\input{content/dataset}

\input{content/evaluation}

\input{content/experiments}
\input{content/results}
\input{content/conclusion_future_work}

\newpage
\bibliography{iclr2026_conference}
\bibliographystyle{iclr2026_conference}

\newpage
\appendix

\input{content/appendix}

\end{document}

%% file: math_commands.tex
\usepackage{amsmath,amsfonts,bm}

\def\eqref#1{equation~\ref{#1}}

\def\1{\bm{1}}

\DeclareMathAlphabet{\mathsfit}{\encodingdefault}{\sfdefault}{m}{sl}
\SetMathAlphabet{\mathsfit}{bold}{\encodingdefault}{\sfdefault}{bx}{n}



%% file: content/introduction.tex
\section{Introduction}
\label{sec:intro}

\begin{figure}[htbp]
    \centering
    
    \begin{subfigure}[t]{0.32\linewidth}
        \centering
        \includegraphics[width=\linewidth]{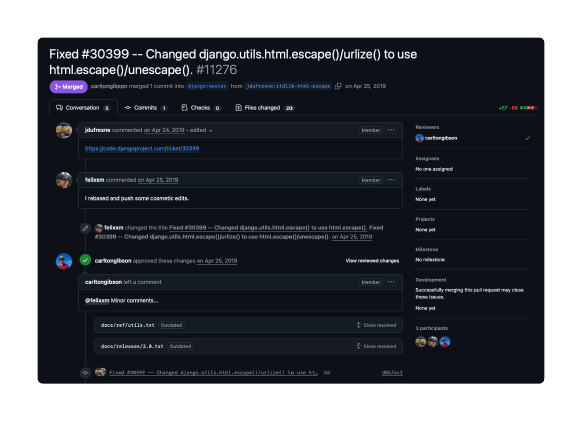}
        \caption{Pull Request ($PR$)}
        \label{fig:pr}
    \end{subfigure}
    \hfill
    \begin{subfigure}[t]{0.32\linewidth}
        \centering
        \includegraphics[width=\linewidth]{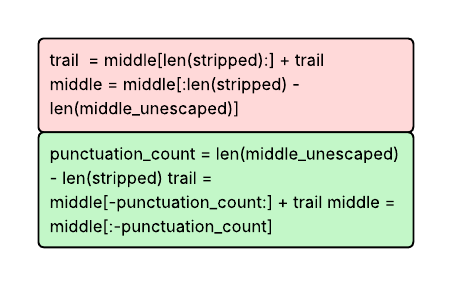}
        \caption{Patch ($r_\delta$)}
        \label{fig:patch}
    \end{subfigure}
    \hfill
    \begin{subfigure}[t]{0.32\linewidth}
        \centering
        \includegraphics[width=\linewidth]{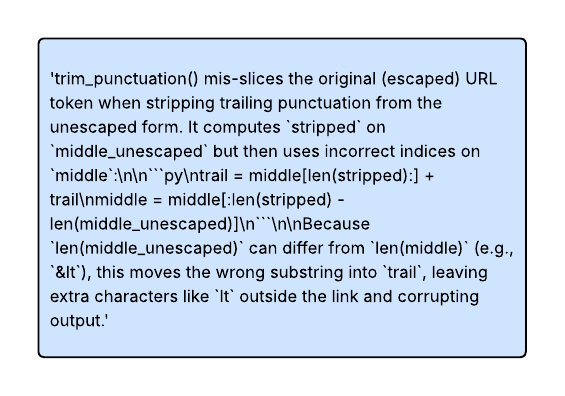}
        \caption{Review Comment ($r_c$)}
        \label{fig:comment}
    \end{subfigure}

    \caption{Example of an instance from \texttt{CR-Bench} dataset: In a pull request, a comment should be made to address the specified bug. The fix is removing the lines of code marked in \textcolor{red}{RED} and adding the ones marked in \textcolor{green}{GREEN}.}
    \label{fig:example}
\end{figure}

Code review is a challenging task to automate~\citep{modern-code-review-survey,modern-code-review-challenges}, as it lacks a well-defined and universally accepted evaluation signal, unlike tasks such as compilation or unit testing where objective pass/fail metrics exist and verification is possible. Further, code review inherently is a complex task due two reasons. First, the code suggestions made during review can be documentation-based, stylistic, refactoring, which are very subjective as per practices followed by different teams. Second, even the comments can be addressed in multiple ways, and there is no uniquely correct way of doing it. 

Due to the complexity of the problem, there develops an inherent bias in code review agents~\citep{coderabbit2024,cursorbugbot2024,googlecodeassist2024,codeagent}. We have observed that these agents often face a fundamental trade-off, they either prioritize precision and risk missing critical vulnerabilities, or prioritize recall at the cost of producing noisy and low-actionable feedback. Gathering feedback of live running agents like Coderabbit~\citep{coderabbit2024}, Gemini code assist~\citep{googlecodeassist2024} from engineers have revealed that more noise in comments hampers developer productivity, inhibiting adoption of agents in real-world code review process. But, avoiding them risks missing defects that might be critical. Consequently, the systematic development and comparison of code review agents necessitates a standardized and principled code review benchmark. 

Existing code review benchmarking datasets often suffer from two primary limitations. First, they frequently combine objective logic errors with subjective stylistic preferences~\citep{swr-bench,harnessing-llm-codereview}. While aspects like style, documentation and formatting are important, they can often better managed by static analysis tools or linting hooks rather than expensive LLM inference. Second, many benchmarks rely on synthetic or small-scale problems~\citep{trans-review-data,t5-codereview,autotransform} that fail to capture the multi-file dependencies and complexities of large-scale repositories. We argue that an effective code review agent must prioritize defect-identifying reviews. These are reviews that target functional, performance, reliability, or security regressions. We argue to focus on this subset for three primary reasons: (1) unaddressed defects directly introduce system faults, and are critical to identify; (2) defect detection is inherently more objective and thus better suited for automated evaluation; and (3) stylistic or structural changes are often subjective and can vary for different projects/teams/working groups.


To address these gaps, we introduce \texttt{CR-Bench} -- a novel benchmarking dataset designed to evaluate the reasoning and defect-detection capabilities of code review agents. In addition, 
we create \texttt{CR-Evaluator}, an evaluation agent that takes a PR, the reviews generated by any black-box agent, and determines its performance, as well as how effective they will be for developers. For PRs in \texttt{CR-Bench}, it is important to identify the underlying bugs. But, a generalized code review agent will also give additional comments, which may, or may not be useful. Therefore, alongside standard metrics like precision, recall, and F1-score of finding a bug, we also introduce two additional metrics -- \textit{usefulness rate} and \textit{signal-to-noise ratio}, which drive measuring utility and developer acceptance of code review agents running in production workflows. This will help in understanding their accuracy, trustworthiness, and factuality. We evaluate two agents using our framework -- one is a single-shot LLM which takes the PR's diff in context and returns reviews on it. Another is a Reflexion~\citep{reflexion} agent that argues and iteratively builds/removes reviews on the PR before making suggestions. Using our experiments, we demonstrate that if we pressure an agent to identify more bugs (like Reflexion), the noise increases, and if we make it too relaxed (like single LM), some bugs are missed. A good agent should fall in a delicate sweet spot between this spectrum.

Our key contributions are as follows: 

\begin{itemize} [leftmargin=*, itemsep=0.1em, topsep=0.1em]
\item \texttt{CR-Bench}, a benchmark for automated code review focusing on real-world, preventable defects, labeled with tags covering bug category, impact, and severity. 
\item \texttt{CR-Evaluator}, a verifier agent that can effectively evaluate a code review agent beyond its accuracy, to trustworthiness, developer acceptability and factuality.
\item Evaluation of two complementary prompting paradigms with two state-of-the-art LLMs to demonstrating performance comparisons and key trade-offs between review coverage and integrity present in code review agents.
\end{itemize}

In the rest of the paper, we discuss these in detail. Section~\ref{sec:related_work} discusses related work, Section~\ref{sec:prob-defn} formulates the problem. Section~\ref{sec:dataset_overview} details the \texttt{CR-Bench} generation methodology, and Section~\ref{sec: cr_verifier} presents the \texttt{CR-Evaluator}. The experiments and results are discussed in Section~\ref{sec:experiments} and~\ref{sec:results} respectively. Finally, Section~\ref{sec:conclusion} concludes the paper and Section~\ref{sec:future_work} discusses future work.

%% file: content/related_work.tex
\section{Related Work}
\label{sec:related_work}

\begin{table*}[t]
    \centering
    \addtolength{\tabcolsep}{-2pt}
    \resizebox{\textwidth}{!}{
    \begin{tabular}{l|cccccc}
        \toprule
        \textbf{Method} & \textbf{\# Tasks} & \textbf{Multiple Diffs} & \textbf{Full PR Context} & \textbf{Categorization} & \textbf{Defect-focused} & \textbf{}{Evaluation} \\
        \midrule
        \citealp{trans-review-data} & $1718$ & \ding{55} & \ding{55} & \ding{55} & \ding{55} & Exact/BLEU \\
        \citealp{t5-codereview} &$ 147,533$ & \ding{55} & \ding{55} & \ding{55} & \ding{55} & Exact/BLEU \\
        \citealp{autotransform} & $~17K$ & \ding{55} & \ding{55} & \ding{55} & \ding{55} & Exact \\
        \citealp{codereviewer} & $54K$ & \ding{55} & \ding{55} & \ding{55} & \ding{55}  & CodeBLEU \\
        \citealp{swr-bench} & $1000$ & \ding{51} & \ding{51} & \ding{51} & \ding{55} & LLM Judge (P, R, F1) \\
        \midrule
        \textbf{CR-Bench (Ours)} & 584 & \ding{51} & \ding{51} & \ding{51} & \ding{51} & LLM Judge (P, R, F1, U, SNR) \\
        \textbf{CR-Bench-Verified (Ours)} & $174$ & \ding{51} & \ding{51} & \ding{51} & \ding{51} & LLM Judge (P, R, F1, U, SNR) \\
        \bottomrule
    \end{tabular}
    }
    \caption{Comparison of Benchmarks for Automated Code Review Evaluation. The data instances in \texttt{CR-Bench} are more defect-focused and real bugs found existing in large-scale open-source repositories. Also, alongside measuring the performance traditionally with precision, recall, F1, we also introduce developer trust and factuality parameters with usefulness score (U), and signal-to-noise ratio (SNR).}
    \small
    \label{tab:datasets-comparison}
\end{table*}

\textbf{Code Review Agents.} Code review agents have evolved from simple code refinement \citep{trans-review-data,t5-codereview} and comment generation \citep{codeagent} to AI agents capable of multi-step reasoning, cross-file interaction, and tool usage. Modern commercial platforms like CodeRabbit~\citep{coderabbit2024}, and Cursor Bugbot~\citep{cursorbugbot2024} are code review products that actively analyze architectural impact, suggest complex changes, and even interact with external project management tools. Similarly, frontier models, including Gemini Code Assist~\citep{googlecodeassist2024} and Claude Code~\citep{claudecode2024} also have their code review tool. Open-source initiatives such as PR-Agent~\citep{qodopragent2024} have further democratized these capabilities. These agents have moved from experimental laboratory tools to mainstream production components. With such wide usage, it is important to evaluate them correctly and determine best available option. This brings a need for good quality benchmark.

\textbf{Automated Code Review Benchmarks.} Early approaches attempted to solve code review by training RNNs~\citep{trans-review-data} and transformers models~\citep{autotransform,t5-codereview} to translate buggy methods into fixed ones. However, these works operated at a restrictive method-level granularity, relying on code abstraction or truncation that omits critical cross-file context. They also evaluated performance using static text-similarity metrics (e.g., BLEU~\citep{bleu}) that fail to capture functional correctness. CodeReviewer~\citep{codereviewer} attempted to specifically pre-train a model optimized for diff-hunks, yet it remains limited to local context and static evaluation. Recently, SWR-Bench~\citep{swr-bench} advanced the field by providing full Pull Request (PR) contexts and evaluating comments using LLM-as-a-judge~\citep{g-eval}. But, it focuses primarily on identifying hit metrics by comparison against human-verified change-points, thus invariably incorporating changes which can be very subjective in nature when compared to objective errors that cause functional errors. In contrast, \texttt{CR-Bench} is the first benchmark to focus only on objective defect-detection with full PR context for agentic evaluation. Table \ref{tab:datasets-comparison} presents a novelty positioning table for \texttt{CR-Bench}. We discuss the formalism of our problem in next section.

%% file: content/problem_definition.tex
\section{Formalism}
\label{sec:prob-defn}

\textbf{Code Review Task Definition.} The \textit{Code Review} is an asynchronous peer-evaluation mechanism aimed at assessing the code quality and recommending possible improvements for a base code state $(B)$. It starts with a \textit{Pull Request} ($PR$) -- a proposal to integrate a sequence of code modifications into a target codebase. Formally, we define a $PR$ as a tuple $(\mathcal{C}, \Delta, \mathcal{M})$, where $\mathcal{C} = \{c_1, c_2, \dots, c_n\}$ is a set of discrete \textit{commits}, $\Delta$ represents the cumulative code diff (patch), and $\mathcal{M}$ denotes the associated metadata (e.g., title, description, and author). Each commit $c \in \mathcal{C}$ constitutes a cryptographic snapshot of the repository state, capturing atomic changes to the directory structure and file contents.
The output for a code review process is a set of $m$ review blocks: $\mathcal{R} = \{ (r_{n}, r_{\delta})_i \mid 1 \le i \le m \}$. $r_{n}$ is a natural language recommendation, which identifies specific defects or suggests improvements, and $r_{\delta}$ provides a concrete code-level resolution to the identified issue. The task of a code review agent is to take a pair of $(PR, B)$ as input and generate a set of $m$ reviews, $\mathcal{R}^{m}$, as shown in Equation \ref{eq:def}.

\begin{equation}
    (PR, B)  \xrightarrow{\text{Code Review Agent}} \mathcal{R}^{m}
    \label{eq:def}
\end{equation}

\textbf{Bias-Variance Trade-off.} Code review tasks,including documentation checks, coding style enforcement, and structural consistency verification, often involve substantial subjectivity~\citep{llm-based-code-review}. In practice, a given change can admit multiple valid review interpretations, and this variability increases with the size and complexity of the underlying repository. Such subjectivity can introduce review noise, making reliable automation challenging. Consequently, designing automated code review agents is a non-trivial task. An overly conservative agent may fail to flag important issues, while an overly permissive agent may generate excessive or low-quality feedback. This reflects a pronounced bias–variance trade-off that represents a central design challenge for automated code review systems.

\textbf{Scope.} This work prioritizes defect-identifying reviews, those targeting functional, performance, reliability, or security regressions. We focus on this subset for three primary reasons: (1) unaddressed logic errors directly introduce system faults; (2) defect detection is inherently more objective and thus better suited for automated evaluation by LLM-based agents; and (3) stylistic or structural preferences are often subjective and better managed via linting tools or human judgment. Following section introduces the dataset to facilitate such study. 

%% file: content/dataset.tex
\section{CR-Bench: Benchmarking Dataset for Code Review}
\label{sec:dataset_overview}

\begin{figure}[htbp]
    \centering
    \includegraphics[width=\linewidth]{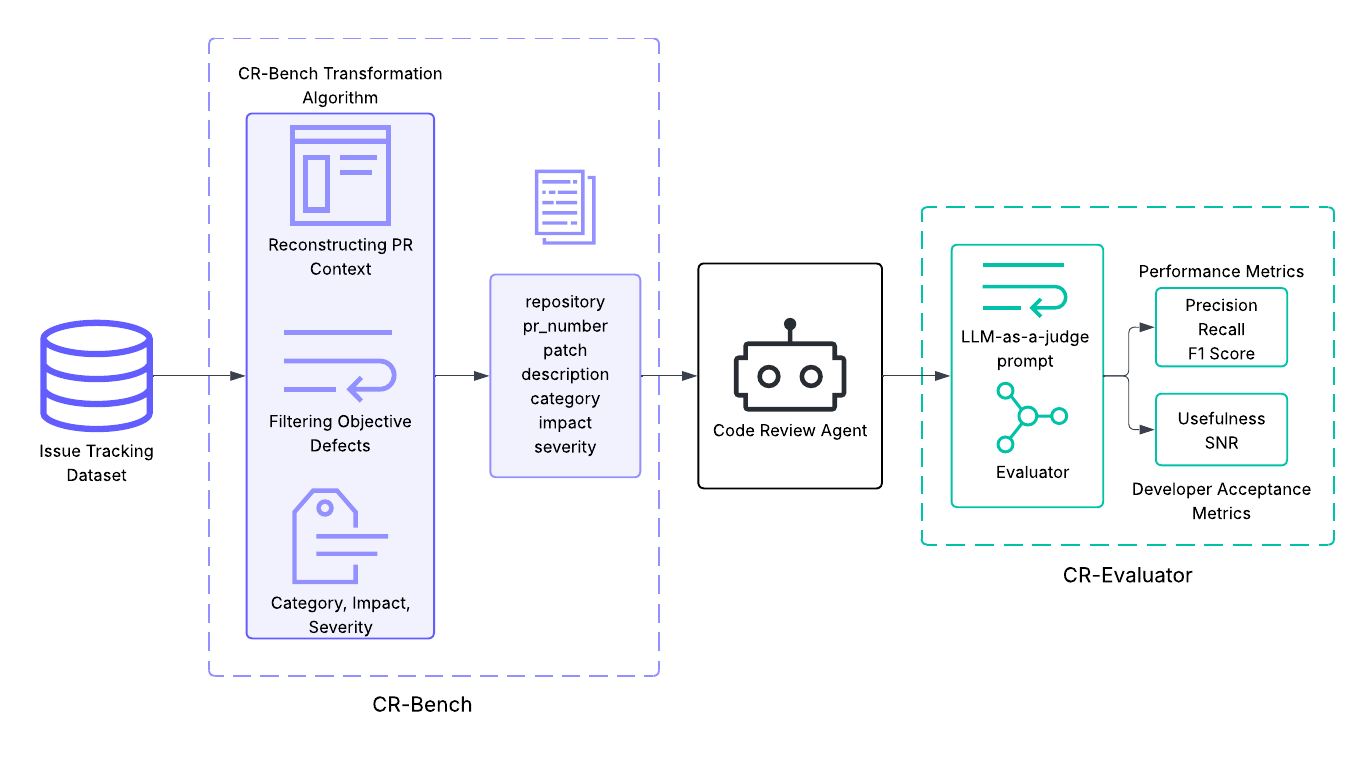}
    \caption{ \texttt{CR-Bench} is a dataset for transforming real-world software defects into an objective code review benchmark with full Pull Request (PR) context and a multi-dimensional taxonomy of Category, Impact, Severity. \texttt{CR-Evaluator} is a method of evaluating code review agents and measuring them in terms of both performance and developer acceptance.}
    \label{fig:intro}
\end{figure}

We offer \texttt{CR-Bench}, a code review dataset $(PR, B) \rightarrow \mathcal{R}^m$. We derived this dataset using SWE-Bench~\citep{swe-bench} as our foundation. SWE-Bench contains real GitHub issues at repository-scale along with the generated PRs that fix them. We tranformed SWE-Bench dataset into a code review dataset using Algorithm \ref{alg:dataset_creation}.


 Unlike SWE-Bench, code review focuses on analyzing pull requests (PRs). A PR is a proposed code change that reviewers must evaluate without knowing in advance where bugs may exist or what type of problems might be introduced. The systematic transformation algorithm to convert SWE-Bench into a code review benchmark, is summarized in Figure~\ref{fig:intro} in addition to Algorithm \ref{alg:dataset_creation}.

During dataset construction, we apply several validation checks to ensure that the bugs included are realistically detectable through code review. We also introduce a taxonomy that labels each instance by bug category, impact, and severity. This allows us to measure how well models and agents perform across different types of bugs and levels of risk. We share two variations -- \texttt{CR-Bench}, which is generated from SWE-Bench, and \texttt{CR-Bench-verified} from SWE-Bench-verified. \texttt{CR-Bench-verified} is also verified by us manually for quality.

\subsection{Transformation of SWE-Bench to CR-Bench}
\label{sec:swe_transformation}

\begin{algorithm}
\caption{CR-Bench Transformation Algorithm}
\label{alg:dataset_creation}
\begin{algorithmic}[1]
\Require SWE-Bench $\mathcal{S}$, LLM $\mathcal{L}$
\Ensure CR-Bench $\mathcal{D}$
\State $\mathcal{D} \gets \emptyset$
\For{each $i \in \mathcal{S}$}
    \State $\mathcal{C} \gets \text{blame}(\text{checkout}(i.commit_{base}))$ \label{alg:dataset_creation-step_blame}
    \State $\mathcal{PR} \gets \text{GitHubAPI}(\mathcal{C})$ \label{alg:dataset_creation-step_api}
    \If{$|\mathcal{PR}| \neq 1$} \label{alg:dataset_creation-step_filter_pr}
        \State \textbf{continue} 
    \EndIf
    \State $\textit{detectable} \gets \mathcal{L}(\mathcal{P^C}(i.patch))$ \label{alg:dataset_creation-step_llm_classify}
    \If{$\neg \textit{detectable}$} \textbf{continue} \label{alg:dataset_creation-step_filter_prev}
    \EndIf
    \State $cmt \gets \text{Paraphrase}(i.desc)$ \label{alg:dataset_creation-step_paraphrase}
    \State $tags \gets \mathcal{L}(P^T(i.desc, i.patch))$ \label{alg:dataset_creation-step_tagging}
    \State $\mathcal{D} \gets \mathcal{D} \cup \{\mathcal{PR}, i.patch, cmt, tags\}$ \label{alg:dataset_creation-step_add}
\EndFor
\State \Return $\mathcal{D}$
\end{algorithmic}
\end{algorithm}

To create \texttt{CR-Bench}, we start with the SWE-Bench dataset and follow Algorithm~\ref{alg:dataset_creation}. For each instance, we first check out to the initial state using the base commit, and execute \texttt{git blame}\footnote{https://git-scm.com/docs/git-blame} on the lines shown removed in SWE Bench's patch. This is to extract all the commit IDs that added those lines (Step~\ref{alg:dataset_creation-step_blame}). Using these commit IDs, we use the GitHub APIs to get the pull request (PR) \footnote{https://docs.github.com/en/rest/commits/commits?apiVersion=2022-11-28\#list-pull-requests-associated-with-a-commit} that was associated with that commit (Step~\ref{alg:dataset_creation-step_api}). If there are multiple PRs, then, the issue couldn't have been identified in a single PR, so, we discard them (Step~\ref{alg:dataset_creation-step_filter_pr}). From here, we have the PR, commit, and diff patch of the code review problem.

Using the remaining task instances, we want to know if the bugs could have been detected during the PR Review. We define detectable bugs as the ones which stem from a defect introduced that can be reasonably identified during the original code review through logic inspection, boundary testing, or adherence to API contracts. This is done by classification using an LLM and prompt $\mathcal{P^C}$(Step~\ref{alg:dataset_creation-step_filter_prev}), and non-preventable bugs are discarded. The result becomes our final set. The prompt for this step is given in Listing~\ref{prompt:detectable_bug} in Appendix~\ref{app:cr_bench}.

As per the definition of code review, along with the PR and patch, we also need a review comment. To obtain this, we paraphrase the problem definition (using prompt listing~\ref{prompt:issue_description_paraphrase}, Appendix~\ref{app:cr_bench}) of the SWE-Bench dataset into a bug description (Step~\ref{alg:dataset_creation-step_paraphrase}). This can either be used directly or further modified by the users to suit their evaluation style and conventions.

Our approach may not provide an exhaustive set of all issues for a given PR. But, we prioritize having high-quality, verified defects that are free from data instances which are falsely included.

\subsection{Taxonomy}

\begin{table}[h]
\centering
\begin{tabular}{llc}
\toprule
\multicolumn{1}{c}{\bf Taxonomy} & \multicolumn{1}{c}{\bf Elements} & \multicolumn{1}{c}{\bf Source} \\
\midrule
Root Cause Category & \begin{tabular}[c]{@{}l@{}}Structural, IIS, RFF, Data, \\ Concurrency, Memory, Security, Coding\end{tabular} & \citep{category-taxonomy-source} \\
\midrule
Impact & \begin{tabular}[c]{@{}l@{}}Functional Suitability, Performance Efficiency, \\ Compatibility, Usability, Reliability, Security, \\ Maintainability, Portability\end{tabular} & ISO Standard \\
\midrule
Severity & Low, Medium, High & --- \\
\bottomrule
\end{tabular}

\caption{The \texttt{CR-Bench} taxonomy elements for classifying code reviews (Note: RFF = Requirements, Features and Functionality, IIS = Interface, Integration and System).}
\label{tab:taxonomy}
\end{table}

For the instances in the candidate set, we tag the reviews into buckets that can enable performance analysis of LLMs and code review agents across these. We create three buckets -- category, impact, and severity. Similar to~\citep{bug-category-DL}, we define category as the root cause of the bug, and the impact as the subsequent effect the bug can potentially cause. We define severity as the extent of impact of the bug on the system. For the category, we use an accepted taxonomy~\citep{category-taxonomy-source} and divide the tasks into -- Structural Bugs, Interface, Integration and System (IIS) Bugs, Requirements, Features and Functionality (RFF) Bugs, Data Bugs, Concurrency Bugs, Memory Bugs, Security Bugs, and Coding Bugs. For impact, we use the ISO/IEC 25010 standard~\citep{ISO25010} and divide it into Functional Suitability, Performance Efficiency, Compatibility, Usability, Reliability, Security, Maintainability, and Portability. Severity is divided into Low, Medium and High.

In order to generate tags for the task instances, we use the problem definition ($desc$) and fix patch with a tag generation prompt $\mathcal{P^T}$ (Listing~\ref{prompt:cateogrize_bugs}, Appendix~\ref{app:cr_bench}) and make an LLM call to get the tags (Step~\ref{alg:dataset_creation-step_tagging}). The prompt also elaborates the definition of each of the items in all three buckets. Statistics of the generated dataset is given in Section~\ref{sec:cr_bench_stats}. The next section describes the evaluation methodology to evaluate any code review agent using this dataset.

%% file: content/evaluation.tex
\section{CR-Evaluator}
\label{sec: cr_verifier}

\texttt{CR-Evaluator} uses an LLM-as-a-judge~\citep{g-eval} approach that employs a zero-shot classification prompt to categorize every review generated by a candidate review model against the historical gold standard defect. \texttt{CR-Evaluator} is presented with the known bug description, the specific files modified in the eventual fix, and the candidate review model's comment. It then performs a discrete classification into one of three categories:

\begin{itemize} [leftmargin=*, itemsep=0.1em, topsep=0.1em]
\item \textbf{Bug Hit.} This category corresponds to reviews that accurately identify or directly relate to the specific logic error described in the ground truth.
\item \textbf{Valid Suggestion.} This category corresponds to reviews with constructive feedback such as stylistic improvements, performance optimizations, or edge-case handling. These reviews are technically sound and significant, but not related to the primary defects in the ground truth set.
\item \textbf{Noise.} This category corresponds to reviews that are factually incorrect, irrelevant to code changes, or insignificant, often indicating a model hallucination.
\end{itemize}

These three categories together form the total generated review. They are mutually exclusive and collectively exhaustive representations of the review space. By aggregating these classifications across the entire corpus, \texttt{CR-Evaluator} generates a multi-dimensional performance profile for any black-box review agent. The first profile metric is \textit{Recall} , which measures the bug coverage rate.

\begin{equation}
\label{eq:recall}
\text{Recall} = \frac{\text{Total Bug Hits}}{\text{Total Bugs}}
\end{equation}

To assess the predictive accuracy of a code review agent, we use {Precision}:
\begin{equation}
\label{eq:precision}
\text{Precision} = \frac{\text{Total Bug Hits}}{\text{Total Reviews}}
\end{equation}

The harmonic mean of these two values provides the overall {F1 Score}:
\begin{equation}
\label{eq:f1}
\text{F1 Score} = 2 \cdot \frac{\text{Precision} \cdot \text{Recall}}{\text{Precision} + \text{Recall}}
\end{equation}

To factor in Valid Suggestions as a dissertate for a code agent, we modify the traditional precision metric and introduce a new metric,  \textit{Usefulness Rate} in Equation \ref{eq:usefulness}:

\begin{equation}
\label{eq:usefulness}
\text{Usefulness Rate} = \frac{\text{Total Bug Hits} + \text{Total Valid Suggestions}}{\text{Total Reviews}}
\end{equation}

Clearly, $\text{Usefulness Rate} > \text{Precision}$. It accounts for the benefits of Code Review agents beyond the bug discovery. 

Among three review categories, Bug Hit and Valid Suggestion clearly measure the beneficial signals, while Noise measure the undesirable part. A good code agent should maximize the $(\text{Bug Hit} + \text{Valid Suggestion})$ count and minimize the Noise count. This relative conflict can measured by \textit{signal-to-noise ratio (SNR)}:

\begin{equation}
\label{eq:snr}
\text{SNR} = \frac{\text{Total Bug Hits} + \text{Total Valid Suggestions}}{\text{Total Noise Count}}
\end{equation}

A high SNR serves as a primary proxy for developer trust by quantifying the ratio of actionable signal to distracting hallucinations. It is a critical diagnostic for identifying agents/model that achieve high recall through high-volume output. A low ratio can trigger developer fatigue and eventual tool abandonment in production workflows.

%% file: content/experiments.tex
\section{Experiments}
\label{sec:experiments}

Using the \texttt{CR-Evaluator}, we evaluate two agents and do a comparative study of their performance. The two agents are:
\begin{enumerate}[leftmargin=*, itemsep=0.1em, topsep=0.1em]
    \item \textbf{Single-shot LM ($A_1$)}: This agent follows a direct zero-shot strategy where the model identifies potential bugs, logic issues, and vulnerabilities in a single pass based on the PR diff and description. It provides a baseline for the model's raw intuition by requiring concise, JSON-formatted feedback on specific file paths and line numbers without further iteration (prompt given in listing~\ref{prompt:single_shot_agent}, Appendix~\ref{app:cr_agents}). This single-shot agent is loosely based on ~\citep{qodopragent2024}.
    \item \textbf{Reflexion Agent ($A_2$)}: Utilizing the Reflexion framework~\citep{reflexion}, this agent performs an initial analysis followed by an iterative self-improvement loop where it is explicitly prompted to discover missed bugs (false negatives) and refine existing comments. This multi-stage approach prioritizes thoroughness and diagnostic accuracy, tasking the agent to re-examine the code to identify overlooked logic errors, security flaws, or resource leaks before finalizing the review (prompt given in listing~\ref{prompt:reflexion_agent}, Appendix~\ref{app:cr_agents}).
\end{enumerate}

We run both the agents using GPT-5.2~\citep{gpt-5,gpt-5.2} and GPT-5-mini~\citep{gpt-5} on \texttt{CR-Bench-verified} version of the dataset, and compare their performance as per metrics defined in section~\ref{sec: cr_verifier}. For verification's LLM-as-a-judge, we use Claude-Sonnet-4.5~\citep{claude-4.5-sonnet} using the prompt detailed in listing~\ref{prompt:cr_verifier} in Appendix~\ref{app:cr_verifier}.

%% file: content/results.tex
\section{Results}
\label{sec:results}

\subsection{CR-Bench Statistics}
\label{sec:cr_bench_stats}

\begin{table}[h]
\centering
\small
\begin{tabular}{lcccc}
\toprule
\textbf{Dataset} & \textbf{\# Lines fixed} & \textbf{\# PR Comments} & \textbf{PR Description Length} \\
\midrule
\texttt{CR-Bench} & 10.28 & 41.03 & 906.63 \\
\texttt{CR-Bench-verified} & 8.69 & 35.83 & 893.59 \\
\bottomrule
\end{tabular}
\caption{Average statistics per instance for \texttt{CR-Bench} and its verified subset. All values represent means calculated across the respective dataset instances.}
\label{tab:crbench_stats}
\end{table}

\begin{figure*}[t]
    \centering
    \begin{subfigure}{\linewidth}
        \centering
        \includegraphics[width=\linewidth]{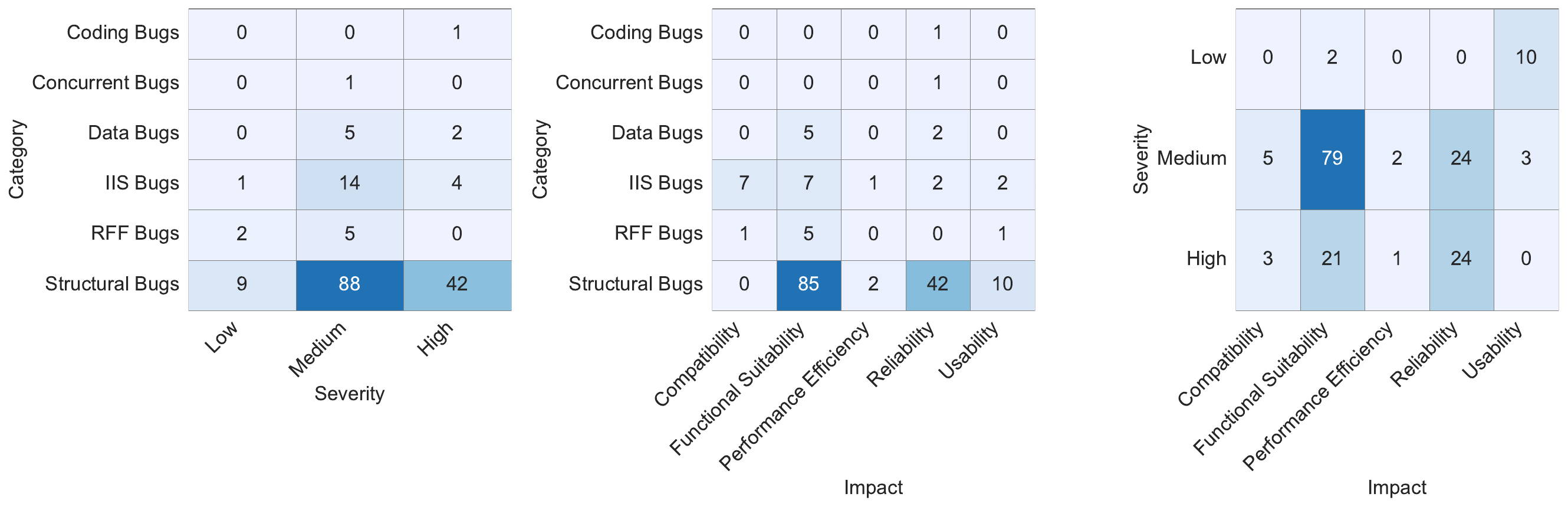}
        \caption{Distribution of category, severity, and impact for the \texttt{CR-Bench-verified} dataset ($N=174$).}
        \label{fig:dist_verified}
    \end{subfigure}
    
    \vspace{1em} 
    
    \begin{subfigure}{\linewidth}
        \centering
        \includegraphics[width=\linewidth]{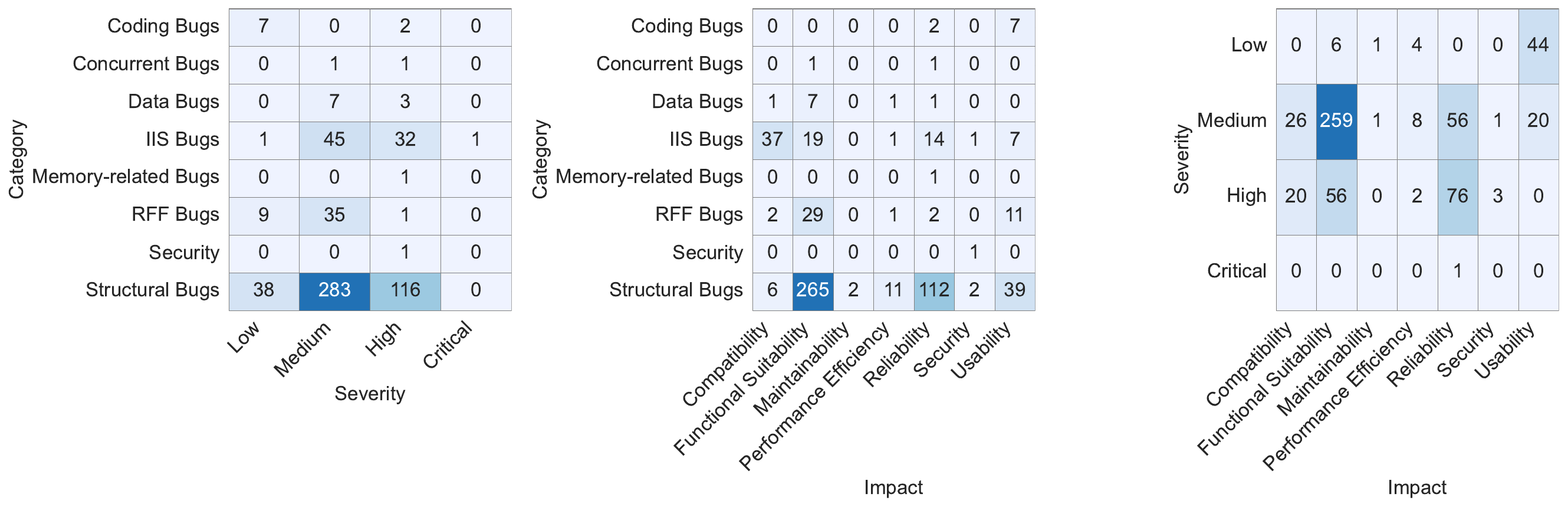}
        \caption{Distribution for the full \texttt{CR-Bench} dataset ($N=584$), showing increased taxonomic diversity.}
        \label{fig:dist_full}
    \end{subfigure}
    
    \caption{Comparative analysis of bug category, severity, and impact distributions between the verified subset and the full \texttt{CR-Bench} corpus (Note: RFF = Requirements, Features and Functionality, IIS = Interface, Integration and System). }
    \label{fig:cr_bench_combined_dist}
\end{figure*}

The final corpus \texttt{CR-Bench-verified} and \texttt{CR-Bench} of consists of $174$ and $584$ high-fidelity PR tasks. The dataset is anchored by major frameworks like django/django, mathematical libraries like sympy/sympy, and specialized tools such as astropy/astropy and scikit-learn/scikit-learn. This shows that these reviews are rooted in mature, long-lived codebases. To summarize the quantitative characteristics of the \texttt{CR-Bench} and \texttt{CR-Bench-verified} datasets, table~\ref{tab:crbench_stats} aggregates the average metrics related to patch complexity, pull request (PR) discussion volume, and context length. These statistics underscore the enterprise-scale nature of the benchmark, highlighting the significant surface area agents must navigate to identify defects.

As per Figure~\ref{fig:dist_verified}, \texttt{CR-Bench-verified} is heavily weighted toward Structural Bugs, accounting for $79.9\%$ of the corpus. These represent complex errors in code arrangement or logic flow that often elude simple static analysis. The primary impacts are Functional Suitability and Reliability. The rigor of \texttt{CR-Bench-verified} is further evidenced by its severity, where $93.1\%$ of the defects are classified as Medium or High severity. High-severity defects are predominantly linked to Reliability and Functional Suitability, representing critical failures that would likely result in system regressions if merged. Thus, while 174 instances may seem less compared to other datasets in Table~\ref{tab:datasets-comparison}, we focus more on quality and stressed evaluation of logical boundaries. While these are quality enforced, there are some categories and impact missing in this subset.

The expansion into the full \texttt{CR-Bench} further reinforces these findings while increasing the statistical power of the benchmark. As shown in Figure~\ref{fig:dist_full}, this larger set maintains the high-stakes nature of the evaluation, with $90.2\%$ of defects classified as Medium, High, or Critical severity. We see instances of impacts like Maintainability, Security existing in this version of the dataset, that were missing before. We also see some minor readjustments to proportions of different categories, impacts, and bugs. This ensures that the benchmark provides a meaningful assessment of an AI agent's ability to safeguard production-grade software.

\subsection{Performance Analysis}
\label{sec:perf}

\begin{table*}[h]
\centering
\small
\begin{tabular}{l|l|ccc|cc}
\toprule
\textbf{Agent} & \textbf{Model} & \textbf{Recall} & \textbf{Prec.} & \textbf{F1} & \textbf{Usefulness} & \textbf{SNR} \\
\midrule
\multirow{2}{*}{Single-shot} & GPT-5.2 & 27.01\% & 3.56\% & 6.30\% & \textbf{83.63\% } & \textbf{5.11} \\
                             & GPT-5-mini & 18.39\% & 3.51\% & 5.90\% & 74.29\% & 2.89 \\
\midrule
\multirow{2}{*}{Reflexion} & GPT-5.2 & \textbf{32.76\%} & \textbf{5.10\%} & \textbf{8.83\%} & 66.10\% & 1.95 \\
                                 & GPT-5-mini & 27.59\% & 3.19\% & 5.72\% & 47.72\% & 0.91 \\
\bottomrule
\end{tabular}
\caption{Comparative performance of code review agents on \texttt{CR-Bench-verified} on two techniques with two models. Best results are marked in \textbf{BOLD}.}
\label{tab:results}
\end{table*}

\textbf{Comparison Across Agents.} The comparative performance of the Single-shot and Reflexion agents illustrates a clear trade-off between discovery coverage and the density of useful feedback. The Single-shot approach demonstrates superior signal integrity, particularly with GPT-5.2. With an SNR of 5.11, it can maintain high developer trust by minimizing the frequency of false alarms. However, this precision comes at the cost of Recall (27.01\%), as the single-pass nature of the agent often overlooks subtle, cross-file errors. Conversely, the Reflexion agents instructively searches for false negatives, which boosted GPT-5.2’s Recall to 32.76\% (an increase from 27.01\%). This technique is most preferable for security-critical audits or high-risk refactoring where catching the needle in the haystack is prioritized over time efficiency. But this comes with a drop of 1.95 in SNR. Moreover, Single-shot agents have lower precision, but higher usefulness compared to Reflexion agents. It could be because of the deep architecture of Reflexion, focusing on bug discovery. 

\textbf{Effect of Model Scale.} The results also highlight a significant divergence in how model scale affects signal stability. GPT-5.2 possesses the semantic robustness to maintain an SNR of $1.95$ even under the pressure of reflexion. In contrast, the SNR of GPT-5-mini agent drops to 0.91, indicating a possible struggle with logical grounding during reflexion. While GPT-5-mini achieves a respectable SNR of 2.89 in a single pass, its SNR collapses to $0.91$ with the reflexion agent. This suggests a Reflexion agent with smaller LLMs  has possibly higher hallucination. Due to iterative probing to identify more issues, the small model possibly suffers from a bias of not being able to disagree that more issues don't exist, and responds with noise.

\subsection{Recall Distribution Across Taxonomy Components}

\begin{figure}[htbp]
     \centering
     \begin{subfigure}[b]{0.32\textwidth}
         \centering
         \includegraphics[width=\textwidth]{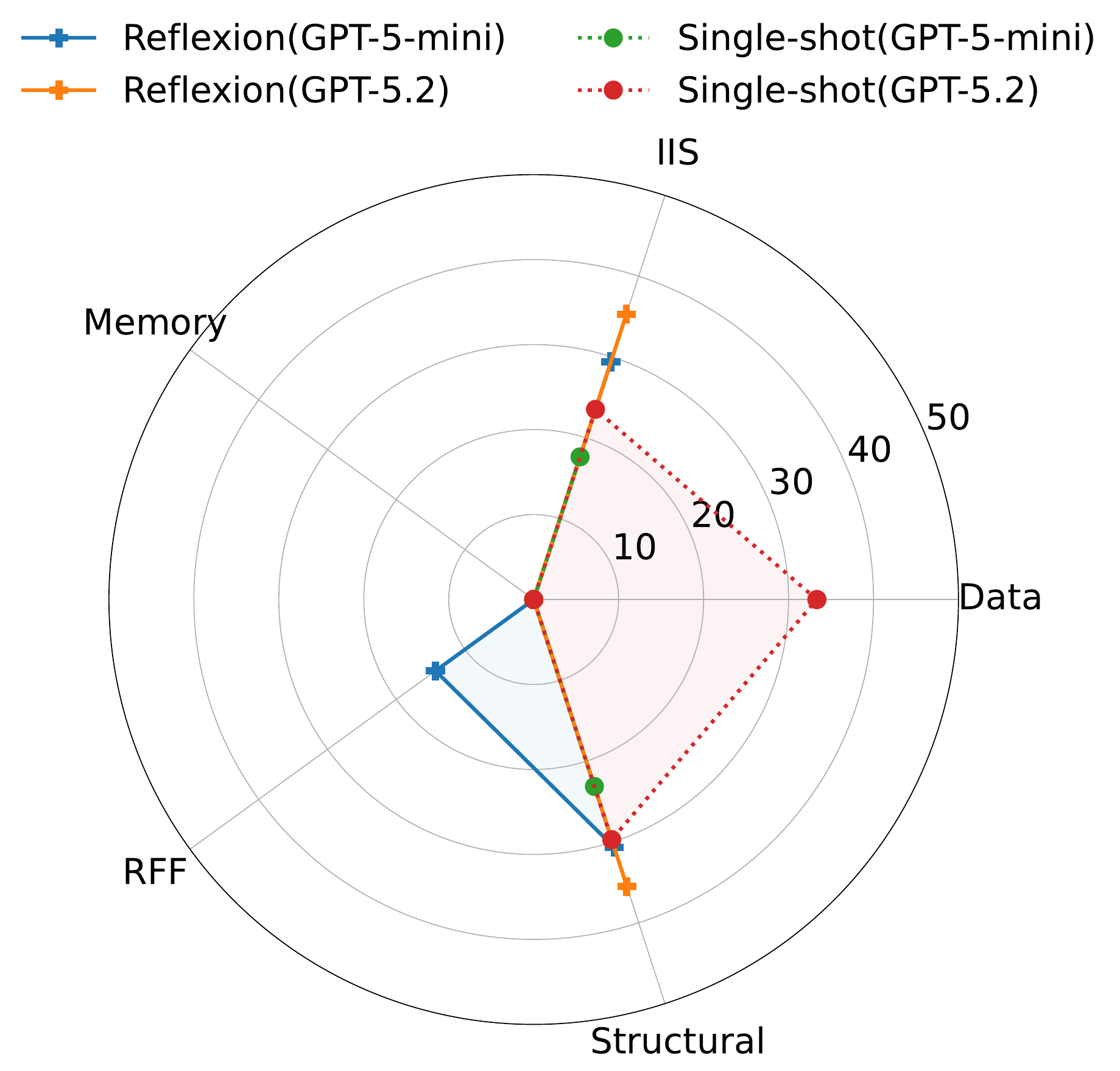}
         \caption{Recall by Category}
         \label{fig:recall_cat}
     \end{subfigure}
     \hfill
     \begin{subfigure}[b]{0.33\textwidth}
         \centering
         \includegraphics[width=\textwidth]{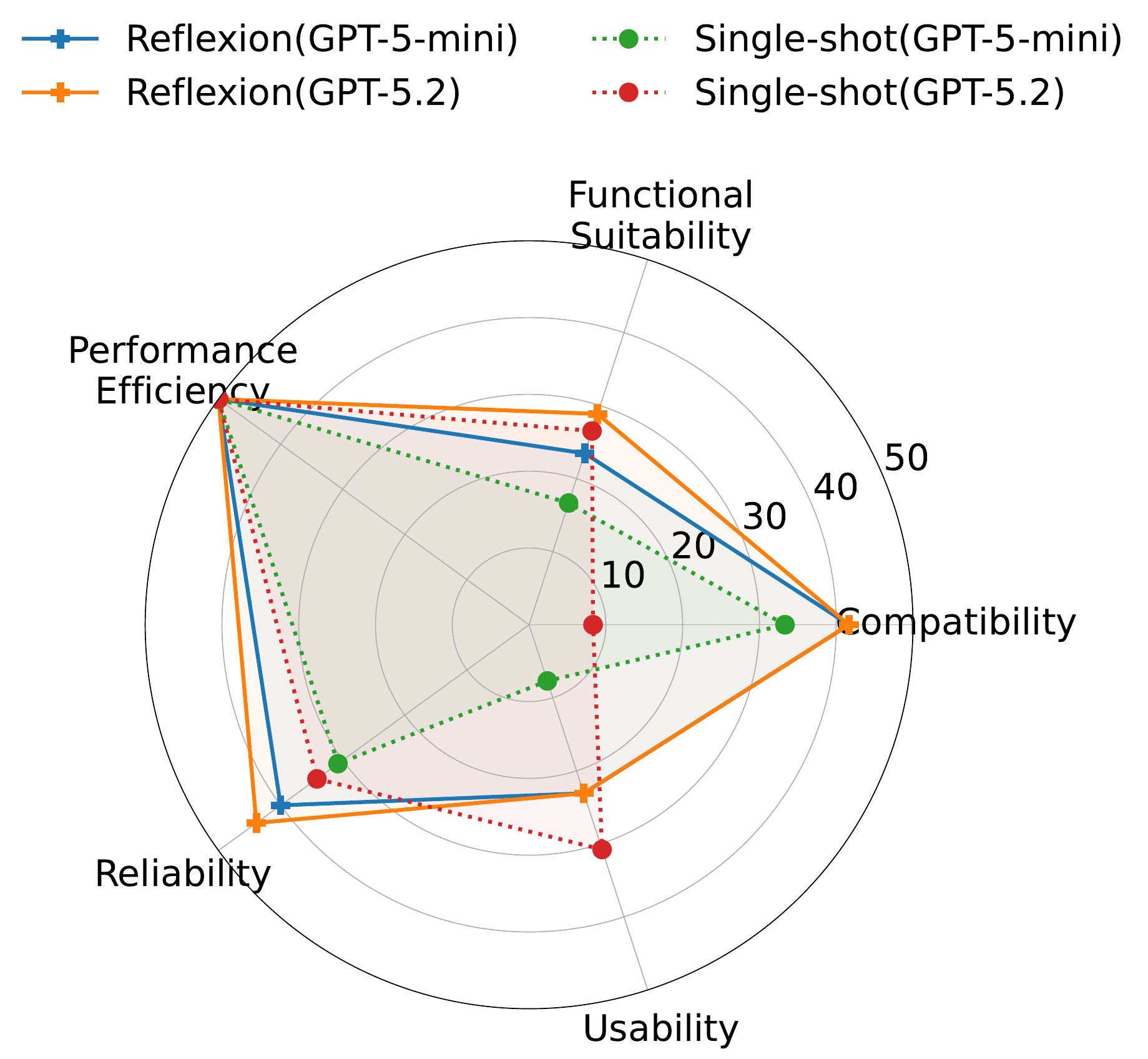}
         \caption{Recall by Impact}
         \label{fig:recall_impact}
     \end{subfigure}
     \hfill
     \begin{subfigure}[b]{0.32\textwidth}
         \centering
         \includegraphics[width=\textwidth]{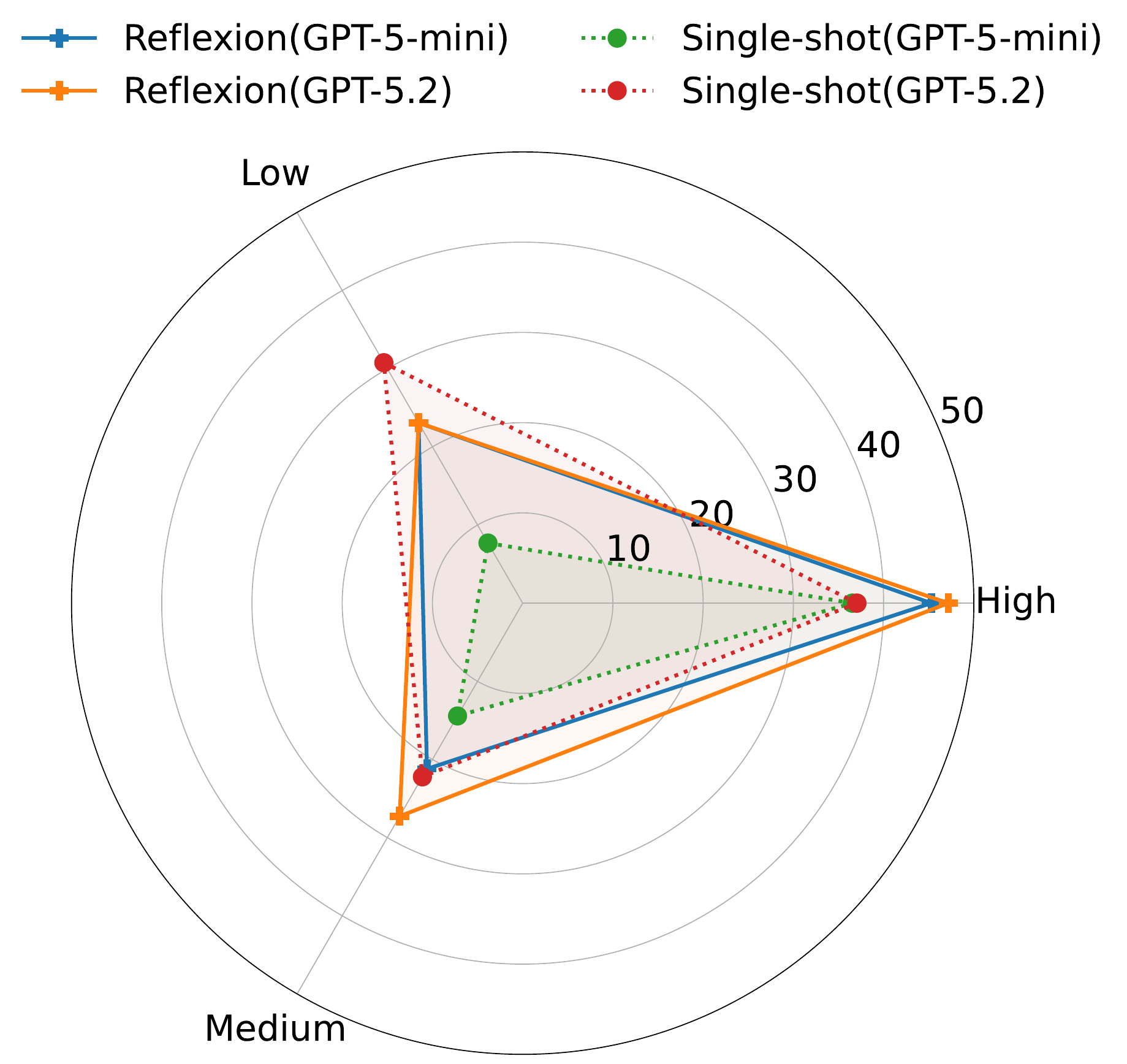}
         \caption{Recall by Severity}
         \label{fig:recall_sev}
     \end{subfigure}
     \caption{PR Bug Recall analysis across Category, Impact, and Severity for \texttt{CR-Bench-verified} (Note: RFF = Requirements, Features and Functionality, IIS = Interface, Integration and System).}
     \label{fig:recall_main}
\end{figure}

Assuming the false negative rate is a leading performance indicator for code review agents, we conducted deep analysis on the recall distribution. 

\textbf{Recall by Root Cause Category. } The distribution of recall across root cause category suggests that all four agents are effective at identifying Structural and IIS (Interface, Integration, and System) defects. Recall of memory issues is zero for all agents, which feels correct as these are determined generally from system traces during execution. The Reflexion framework provides its most significant gains in the Structural and Data categories. For GPT-5.2, the iterative reflexion process allows it to re-read complex logic flows, successfully uncovering deeper algorithmic errors that were missed in the initial heuristic pass, as evidenced by the expanded radar profile in Figure~\ref{fig:recall_cat}. Conversely, GPT-5-mini exhibits a severe performance ceiling in RFF, IIS, and Structural bugs, but with reflexion. The performance degrades with single-shot. A surprising behavior is for Data bugs, where only the performance of single-shot GPT 5.2 is acceptable.

\textbf{Recall by Possible Downstream Impact. } When mapped to ISO/IEC 25010 impact standards, a clear pattern emerges regarding the type of quality an AI agent can protect. All configurations show peak performance in identifying defects with Performance efficiency and Reliability impacts. These bugs often have distinct code signatures such as inefficient loops or missing error handlers, that align well with the pattern-matching strengths of frontier LLMs like GPT-5.2. However, there is a dip in catching Usability and Functional suitability issues. These defects often require external environmental knowledge or upstream dependency context that is not fully contained within the PR diff, highlighting a fundamental limitation in current closed-context code review agents, lacking access to the broader system state. On compatibility issues, all agents except single-shot GPT 5.2 perform well.

\textbf{Recall by Operational Severity.} On operational severity metrics, all agents demonstrate significantly higher recall for High-Severity defects compared to Low or Medium ones. The GPT-5.2 Reflexion agent achieves its highest discovery rate in the High-severity bucket, suggesting the need for reasoning depth for severe bugs. Conversely, Low-Severity defects show the lowest recall across all paradigms. This indicates that while agents are effective safety nets for major regressions, they are less reliable at identifying minor nit-pick logic flaws, which developers might find acceptable.

%% file: content/conclusion_future_work.tex
\section{Conclusion}
\label{sec:conclusion}

In this work, we introduced \texttt{CR-Bench}, a benechmark dataset designed to bridge the gap between synthetic benchmarks and the complexities of real-world automated code review. By transforming high-fidelity software failures from SWE-Bench into \texttt{CR-Bench}, we provided a rigorous, defect-identifying PR review corpus of $174$ (verified) and $584$ (standard) defects that challenges AI agents to perform blind audits without prior defect knowledge. Our development of \texttt{CR-Evaluator} further enables a nuanced assessment of agent performance, moving beyond binary traditional metrics like precision, recall, accuracy, to more practical metrics from a developers perspective like usefulness rate and signal-to-noise ratio (SNR).

Our systematic evaluation of code review agents uncovered a fundamental design trade-off. While the Reflexion paradigm significantly enhances discovery Recall, it simultaneously incurs a steep cost in signal integrity with high SNR. This effect is most pronounced in lightweight models like GPT-5-mini. Overall, \texttt{CR-Bench} and \texttt{CR-Evaluator} jointly lay the design foundation of real-world code review agent evaluation.

\section{Future Work}
\label{sec:future_work}

The findings from our evaluation of \texttt{CR-Bench} provide a clear roadmap for the next generation of automated code review agents. In this work, we considered four LLM-based agents. In future work, we plan to evaluate a broader range of agents, encompassing additional LLMs and diverse agentic architectures. We will also explore the applicability of advanced post-training techniques, such as GRPO. Finally, we aim to broaden the scope of \texttt{CR-Bench} by incorporating additional programming languages and domain-specific frameworks. This expansion will allow us to investigate how language-specific constraints influence an agent’s signal-to-noise ratio (SNR) and recall profiles.

%% file: content/appendix.tex
\section{Appendix: CR-Bench Prompts}
\label{app:cr_bench}

\begin{lstlisting}[
    basicstyle=\small\ttfamily,
    breaklines=true,
    frame=tb,
    caption={Prompt for classifying a SWE Bench bug as detectable/not detectable.},
    label={prompt:detectable_bug},
    captionpos=b
]
system = """You are a software engineering expert analyzing code changes.
Your task is to determine if a change is of the kind that might have been prevented by reviewing the PR/commit that made that change more carefully or NOT.
Name these PREVENTABLE and UNPREVENTABLE.
PREVENTABLE: Code changes that fix incorrect behavior, errors, crashes, or unexpected results introduced by a previous code change (PR/commit).
UNPREVENTABLE: Everything else including but not limited to: New features, refactoring, documentation, performance improvements, or enhancements.

Respond with JSON only:
{ "category": "PREVENTABLE" or "UNPREVENTABLE", "confidence": 0.0 to 1.0, "reasoning": "brief explanation" }"""

user = """Problem Statement:
{problem_statement}

Patch:
{patch}

Is this a bug fix that might have been introduced by a previous code change (PR/commit)?"""
\end{lstlisting}

\begin{lstlisting}[
    basicstyle=\small\ttfamily,
    breaklines=true,
    frame=tb,
    caption={Prompt for paraphrasing the issue description.},
    label={prompt:issue_description_paraphrase},
    captionpos=b
]
system="""You are a software engineering expert writing technical bug descriptions.

Given a problem statement and the code fix patch, write a concise technical description of the bug.

**Format Requirements:**
- Start with a clear statement of what's wrong
- Include specific code snippets showing the problematic pattern
- Explain why it's a bug (runtime impact, incorrect behavior)
- Keep it under 100 words
- Use technical language
- Focus on the root cause, not symptoms

Respond with ONLY the bug description (no JSON, no extra formatting)."""

user="""Bug Description Task

## Problem Statement
{row['problem_statement']}

## Code Fix
File: {file_path}
Lines: {line_start}-{line_end}

```diff
{patch}
```

Write a concise technical description of this bug."""
\end{lstlisting}

\begin{lstlisting}[
    basicstyle=\small\ttfamily,
    breaklines=true,
    frame=tb,
    caption={Prompt for cateogorizing PR and bugs into category, impact, severity.},
    label={prompt:cateogrize_bugs},
    captionpos=b
]
system = """You are a software engineering expert analyzing bug reports to classify them according to a predefined bug taxonomy used in empirical software engineering research.

Analyze the bug from the issue description (problem statement) and the code fix patch, and provide a classification using the following fields.

---

### 1. **category**

Choose EXACTLY ONE of the following bug types:

#### **1. Requirements, Features, and Functionality Bugs**

**Source:** Ambiguous, incomplete, misunderstood, or changing requirements/specifications.
**Types:**

* **Requirements bugs:** Missing, ambiguous, or inconsistent requirements.
* **Feature bugs:** Wrong, missing, or extra features; undesired enhancements.
  **Detection/Remedy:** Formal specification languages, functional testing (transaction flow, domain, logic, state-based testing).

#### **2. Structural Bugs**

**Source:** Code control flow, sequence, and logical structure.
**Types:**

* **Control & sequence bugs:** Missing paths, unreachable code, improper loop nesting, misused GOTO/labels.
* **Logic bugs:** Incorrect logical expressions; Boolean errors.
* **Processing bugs:** Arithmetic, algorithm, type conversion, overflow errors.
* **Initialization bugs:** Uninitialized variables or improper first-loop values.
* **Data-flow anomalies:** Using uninitialized or stale data; not storing/modifying data correctly.
  **Detection/Remedy:** Structural/path testing, unit testing, domain testing.

#### **3. Data Bugs**

**Source:** Specification, format, initial values, and usage of data objects.
**Types:**

* **Dynamic data bugs:** Temporary data issues, shared memory residues.
* **Static data bugs:** Fixed content/parameters; preprocessing errors.
* **Role-based bugs:** Errors in data used as control, parameter, or information.
* **Attributes, structure, content bugs:** Incorrect semantics, memory allocation, or representation.
  **Detection/Remedy:** Validate all data declarations, centralize shared resources, defensive coding, compile-time checks.

#### **4. Coding Bugs**

**Source:** Programming errors, including wild-card or arbitrary mistakes.
**Types:**

* **Syntax errors:** Usually caught by compilers.
* **Documentation bugs:** Misleading comments or manuals.
  **Detection/Remedy:** Proper code review, accurate documentation, automated syntax checking.

#### **5. Interface, Integration, and System Bugs**

**Source:** Interactions between components, systems, or hardware.
**Types:**

* **External interface bugs:** Wrong device or protocol handling.
* **Internal interface bugs:** Improper module communication.
* **Hardware bugs:** Misunderstood device behavior, I/O errors.
* **OS bugs:** Misuse or assumption of OS services.
* **Software architecture bugs:** Emergent behavior due to module interactions or stress.
* **Integration bugs:** Incompatibilities between integrated modules.
* **System bugs:** Complex interaction of multiple components, rare but expensive.
  **Detection/Remedy:** Design for modularity, stress testing, integration testing, specialist interfaces.

#### **6. Test and Test Design Bugs**

**Source:** Errors in testing processes or test code.
**Types:**

* **Test execution bugs:** Incorrect test implementation.
* **Test design bugs:** Flawed test scenarios or criteria.
  **Detection/Remedy:** Test debugging, automation of test execution and design, test quality assurance.

#### **7. Memory-related Bugs**

**Source:** Improper memory handling.
**Types:**

* **Buffer overflow:** Access beyond allocated memory.
* **Stack smashing:** Overwriting function return addresses.
* **Memory leak:** Lost pointers to allocated memory.
* **Uninitialized read:** Accessing memory before initialization.
* **Double free:** Freeing memory twice.
  **Detection/Remedy:** Careful memory management, automated memory checking tools.

#### **8. Concurrent Bugs**

**Source:** Multi-threading or multi-process synchronization issues.
**Types:**

* **Data race bugs:** Conflicting access to shared memory.
* **Atomicity bugs:** Interrupted sequences of operations.
* **Deadlock:** Processes waiting indefinitely for shared resources.
  **Detection/Remedy:** Proper synchronization, lock ordering, deadlock detection mechanisms.

---

### 2. **severity**

Classify the runtime impact of the bug:

* **Critical:** System crash, severe malfunction, data loss, or serious security risk.
* **High:** Core functionality broken or system unusable in common scenarios.
* **Medium:** Partial malfunction, degraded behavior, or non-critical failures.
* **Low:** Minor issues, cosmetic problems, edge cases, or test-only failures.

---

### 3. **impact**

Choose EXACTLY ONE aspect of the software that the bug directly compromises:

#### **1. Functional Suitability**

* **Description:** Bugs that cause incorrect, missing, or inappropriate behavior relative to requirements.
* **Examples:** Wrong outputs, missing features, incorrect calculations.

#### **2. Performance Efficiency**

* **Description:** Bugs that degrade system performance or resource utilization.
* **Examples:** Slow response, high CPU/memory usage, poor throughput.

#### **3. Compatibility**

* **Description:** Bugs that interfere with interoperability or co-existence with other systems, devices, or software.
* **Examples:** Integration failures, protocol mismatches, API incompatibilities.

#### **4. Usability**

* **Description:** Bugs that make the system confusing, hard to use, or increase user errors.
* **Examples:** Misleading error messages, poor workflow, inaccessible UI elements.

#### **5. Reliability**

* **Description:** Bugs that cause system instability, crashes, incorrect state, or inability to recover from failures.
* **Examples:** Crashes, deadlocks, memory leaks, data corruption.

#### **6. Security**

* **Description:** Bugs that expose the system to threats, unauthorized access, or data compromise.
* **Examples:** Buffer overflows, improper access control, injection vulnerabilities.

#### **7. Maintainability**

* **Description:** Bugs that make code harder to understand, modify, or extend.
* **Examples:** Poorly structured code, missing documentation, complex interdependencies.

#### **8. Portability**

* **Description:** Bugs that prevent software from being installed, executed, or transferred across different environments.
* **Examples:** OS-specific assumptions, hardware incompatibilities, environment-specific failures.

---

### 4. **reasoning**

Concisely explain the reasoning behind the **category**, **impact**, and **severity** of the bug.

---

### Output format

Respond ONLY with valid JSON in the following format:

```json
{
  "category": "...",
  "severity": "...",
  "impact": "...",
  "reasoning": "..."
}
```

Do not include explanations, markdown, or any additional text outside the JSON."""

user = """# Bug Classification Task

## Problem Statement
{problem_statement}

## Code Fix
File: {file_path}
Lines: {line_start}-{line_end}

```diff
{correct_fix}
```

Analyze this bug and provide classification as JSON."""
\end{lstlisting}

\section{Appendix: Code Review Agent Prompts}
\label{app:cr_agents}

\begin{lstlisting}[
    basicstyle=\small\ttfamily,
    breaklines=true,
    frame=tb,
    caption={Prompt for Single-shot agent.},
    label={prompt:single_shot_agent},
    captionpos=b
]
SYSTEM_PROMPT = """You are an expert code reviewer analyzing pull requests for potential bugs and issues.

Your task is to review the provided code changes and identify:
1. Potential bugs or errors
2. Logic issues
3. Edge cases not handled
4. Security vulnerabilities
5. Performance problems

For each issue found, provide:
- file: The file path
- line: The line number (approximate)
- comment: Clear description of the issue
- severity: "high", "medium", or "low"

Output your review as a JSON array of comments. Example:
[
 {
   "file": "src/module.py",
   "line": 42,
   "comment": "Potential null pointer dereference. Variable 'user' may be None.",
   "severity": "high"
 }
]

If no issues found, return an empty array: []

Focus on actual bugs, not style or formatting issues. Do not be verbose, just provide the review comments in concise manner that descirbes the issue completely."""

USER_PROMPT_TEMPLATE = Template("""# Pull Request Review

**Repository:** {{ repo }}
**PR Number:** #{{ pr_number }}
**Title:** {{ title }}
**Description:** {{ description }}

## Diff
```diff
{{ diff }}
```

Please review the above changes and identify any potential bugs or issues.""")
\end{lstlisting}

\begin{lstlisting}[
    basicstyle=\small\ttfamily,
    breaklines=true,
    frame=tb,
    caption={Prompt for Reflexion agent.},
    label={prompt:reflexion_agent},
    captionpos=b
]
SYSTEM_PROMPT = """You are an expert code reviewer specializing in bug detection and security analysis.

Review the provided code changes and identify ONLY actual bugs and issues, NOT style or formatting concerns.

## Focus Areas:
1. **Bugs**: Logic errors, incorrect implementations, type mismatches
2. **Edge Cases**: Unhandled null/undefined, empty arrays, boundary conditions
3. **Security**: SQL injection, XSS, authentication bypasses, data leaks
4. **Concurrency**: Race conditions, deadlocks, thread safety issues
5. **Performance**: Memory leaks, infinite loops, inefficient algorithms

## Output Format:
Return a JSON array. Each issue must include:
- file: Exact file path from the diff
- line: Line number where issue occurs
- comment: Specific, actionable description of the bug (not generic warnings)
- severity: "high" (crashes/security), "medium" (data corruption/incorrect behavior), "low" (minor issues)

## Quality Standards:
- Be SPECIFIC: Point to exact problematic code, not general concerns
- Be CERTAIN: Only report issues you're confident are real bugs
- Explain WHY: Include the consequence of the bug
- Suggest FIX: Briefly mention how to resolve it

Return [] if no bugs found.

Example:
[
  {
    "file": "src/auth.py",
    "line": 23,
    "comment": "Missing null check before accessing 'user.email'. Will throw AttributeError if user is None. Add: if user is None: return error",
    "severity": "high"
  }
]"""

USER_PROMPT_TEMPLATE = Template("""# Pull Request to Review

**Repository:** {{ repo }}
**PR #{{ pr_number }}:** {{ title }}

{{ description }}

## Code Changes
```diff
{{ diff }}
```

Analyze the diff above and identify concrete bugs. Focus on what could actually break or cause incorrect behavior.""")

REFLECTION_SYSTEM_PROMPT = """You are performing iterative self-improvement using the Reflexion framework.

Your goal: Produce a MORE COMPLETE and ACCURATE review by finding bugs you missed.

## Reflexion Strategy:

### PRIMARY FOCUS: Find Missed Bugs (False Negatives)
Your previous review likely MISSED important bugs. Search for:
- Null/undefined dereferences you overlooked
- Logic errors in conditionals or loops
- Off-by-one errors, incorrect operators
- Unhandled exceptions or error cases
- Security vulnerabilities (injection, auth bypass)
- Resource leaks, race conditions, memory issues

### SECONDARY: Remove Only OBVIOUS False Positives

**When in doubt, KEEP the issue.** Err on the side of caution.

### TERTIARY: Improve Existing Issues
- Make vague comments more specific
- Adjust severity if clearly wrong
- Add fix suggestions

## Output Requirements:
- Return the FULL JSON array with ALL valid issues
- Prioritize ADDING missed bugs over removing uncertain ones
- Each comment must point to specific problematic code
- Be thorough - it's better to report a potential bug than miss a real one

Return [] only if you're absolutely certain no bugs exist."""

REFLECTION_USER_PROMPT_TEMPLATE = Template("""# Reflexion - Iteration {{ iteration }}: Find What You Missed

**Repository:** {{ repo }} | **PR:** #{{ pr_number }}

## Code to Review
```diff
{{ diff }}
```

---

## Your Previous Review (Iteration {{ prev_iteration }})
You found {{ num_comments }} issue(s):
```json
{{ previous_comments }}
```

---

## Reflexion Task: FIND MISSED BUGS

### PRIMARY: What Bugs Did You Miss?

Carefully re-examine the diff. Look for bugs you 

## Output Your Complete Refined Review

Return a JSON array with:
1. All valid issues from previous review (unless clearly false positive)
2. NEW bugs you discovered in this reflection
3. Improved comments and severity levels

Format:
```json
[
  {
    "file": "path/to/file.py",
    "line": 42,
    "comment": "Specific bug: what breaks, why, and how to fix",
    "severity": "high|medium|low"
  }
]
```

Return [] only if absolutely no bugs exist.""")
\end{lstlisting}

\section{Appendix: CR-Evaluator Prompts}
\label{app:cr_verifier}

\begin{lstlisting}[
    basicstyle=\small\ttfamily,
    breaklines=true,
    frame=tb,
    caption={Prompt for CR-Evaluator.},
    label={prompt:cr_verifier},
    captionpos=b
]
"""You are evaluating code review comments against a known bug.

## Known Bug Description:
{bug_description}

## Files Changed in Fix:
{patch_files}

## Review Comment to Evaluate:
File: {file}
Line: {line}
Comment: {comment}

## Task:
Classify this review comment into ONE of these categories:

1. **BUG_HIT**: The comment identifies or relates to the same issue described in the bug description
2. **VALID_SUGGESTION**: The comment makes a valid point (style, performance, maintainability, edge case, etc.) but is NOT about the known bug
3. **NOISE**: The comment is incorrect, irrelevant, or not actionable

Respond with ONLY a JSON object:
{{"classification": "BUG_HIT|VALID_SUGGESTION|NOISE", "reason": "brief explanation"}}
"""
\end{lstlisting}

%% file: iclr2026_conference.bib
@misc{modern-code-review-survey,
      title={A Survey on Modern Code Review: Progresses, Challenges and Opportunities}, 
      author={Zezhou Yang and Cuiyun Gao and Zhaoqiang Guo and Zhenhao Li and Kui Liu and Xin Xia and Yuming Zhou},
      year={2024},
      eprint={2405.18216},
      archivePrefix={arXiv},
      primaryClass={cs.SE},
      url={https://arxiv.org/abs/2405.18216}, 
}

@inproceedings{codeagent,
    title = "{C}ode{A}gent: Autonomous Communicative Agents for Code Review",
    author = "Tang, Xunzhu  and
      Kim, Kisub  and
      Song, Yewei  and
      Lothritz, Cedric  and
      Li, Bei  and
      Ezzini, Saad  and
      Tian, Haoye  and
      Klein, Jacques  and
      Bissyand{\'e}, Tegawend{\'e} F.",
    editor = "Al-Onaizan, Yaser  and
      Bansal, Mohit  and
      Chen, Yun-Nung",
    booktitle = "Proceedings of the 2024 Conference on Empirical Methods in Natural Language Processing",
    month = nov,
    year = "2024",
    address = "Miami, Florida, USA",
    publisher = "Association for Computational Linguistics",
    url = "https://aclanthology.org/2024.emnlp-main.632/",
    doi = "10.18653/v1/2024.emnlp-main.632",
    pages = "11279--11313"
}

@misc{harnessing-llm-codereview,
      title={Harnessing Large Language Models for Curated Code Reviews}, 
      author={Oussama Ben Sghaier and Martin Weyssow and Houari Sahraoui},
      year={2025},
      eprint={2502.03425},
      archivePrefix={arXiv},
      primaryClass={cs.SE},
      url={https://arxiv.org/abs/2502.03425}, 
}

@INPROCEEDINGS{modern-code-review-challenges,
  author={Bacchelli, Alberto and Bird, Christian},
  booktitle={2013 35th International Conference on Software Engineering (ICSE)}, 
  title={Expectations, outcomes, and challenges of modern code review}, 
  year={2013},
  volume={},
  number={},
  pages={712-721},
  doi={10.1109/ICSE.2013.6606617}}

@misc{llm-based-code-review,
      title={Benchmarking and Studying the LLM-based Code Review}, 
      author={Zhengran Zeng and Ruikai Shi and Keke Han and Yixin Li and Kaicheng Sun and Yidong Wang and Zhuohao Yu and Rui Xie and Wei Ye and Shikun Zhang},
      year={2025},
      eprint={2509.01494},
      archivePrefix={arXiv},
      primaryClass={cs.SE},
      url={https://arxiv.org/abs/2509.01494}, 
}

@misc{coderabbit2024,
  author = {{CodeRabbit}},
  title = {CodeRabbit: AI-powered Code Reviews},
  year = {2024},
  url = {https://www.coderabbit.ai/},
  note = {Accessed: 2024-05-20}
}

@misc{cursorbugbot2024,
  author = {{Anysphere}},
  title = {Cursor Bugbot: Automated Bug Finding and Fixing},
  year = {2024},
  url = {https://cursor.com/bugbot},
  note = {Accessed: 2024-05-20}
}

@misc{googlecodeassist2024,
  author = {{Google Cloud}},
  title = {Gemini Code Assist: AI-powered assistance for software development},
  year = {2024},
  url = {https://codeassist.google/},
  note = {Accessed: 2024-05-20}
}

@misc{claudecode2024,
  author = {{Anthropic}},
  title = {Claude Code: GitHub Actions Documentation},
  year = {2024},
  url = {https://code.claude.com/docs/en/github-actions},
  note = {Accessed: 2024-05-20}
}

@misc{qodopragent2024,
  author = {{Qodo AI}},
  title = {PR-Agent: AI-powered tool for automated pull request analysis, feedback, and suggestions},
  year = {2024},
  publisher = {GitHub},
  journal = {GitHub repository},
  url = {https://github.com/qodo-ai/pr-agent},
  note = {Accessed: 2024-05-20}
}

@inproceedings{trans-review-data,
author = {Tufano, Rosalia and Pascarella, Luca and Tufano, Michele and Poshyvanyk, Denys and Bavota, Gabriele},
title = {Towards Automating Code Review Activities},
year = {2021},
isbn = {9781450390859},
publisher = {IEEE Press},
url = {https://doi.org/10.1109/ICSE43902.2021.00027},
doi = {10.1109/ICSE43902.2021.00027},
booktitle = {Proceedings of the 43rd International Conference on Software Engineering},
pages = {163–174},
numpages = {12},
location = {Madrid, Spain},
series = {ICSE '21}
}

@inproceedings{autotransform,
author = {Thongtanunam, Patanamon and Pornprasit, Chanathip and Tantithamthavorn, Chakkrit},
title = {AutoTransform: automated code transformation to support modern code review process},
year = {2022},
isbn = {9781450392211},
publisher = {Association for Computing Machinery},
address = {New York, NY, USA},
url = {https://doi.org/10.1145/3510003.3510067},
doi = {10.1145/3510003.3510067},
booktitle = {Proceedings of the 44th International Conference on Software Engineering},
pages = {237–248},
numpages = {12},
location = {Pittsburgh, Pennsylvania},
series = {ICSE '22}
}

@inproceedings{t5-codereview,
author = {Tufano, Rosalia and Masiero, Simone and Mastropaolo, Antonio and Pascarella, Luca and Poshyvanyk, Denys and Bavota, Gabriele},
title = {Using pre-trained models to boost code review automation},
year = {2022},
isbn = {9781450392211},
publisher = {Association for Computing Machinery},
address = {New York, NY, USA},
url = {https://doi.org/10.1145/3510003.3510621},
doi = {10.1145/3510003.3510621},
booktitle = {Proceedings of the 44th International Conference on Software Engineering},
pages = {2291–2302},
numpages = {12},
location = {Pittsburgh, Pennsylvania},
series = {ICSE '22}
}

@article{codereviewer,
  title={CodeReviewer: Pre-Training for Automating Code Review Activities},
  author={Li, Zhiyu and Lu, Shuai and Guo, Daya and Duan, Nan and Jannu, Shailesh and Jenks, Grant and Majumder, Deep and Green, Jared and Svyatkovskiy, Alexey and Fu, Shengyu and others},
  journal={arXiv preprint arXiv:2203.09095},
  year={2022}
}

@misc{swr-bench,
      title={Benchmarking and Studying the LLM-based Code Review}, 
      author={Zhengran Zeng and Ruikai Shi and Keke Han and Yixin Li and Kaicheng Sun and Yidong Wang and Zhuohao Yu and Rui Xie and Wei Ye and Shikun Zhang},
      year={2025},
      eprint={2509.01494},
      archivePrefix={arXiv},
      primaryClass={cs.SE},
      url={https://arxiv.org/abs/2509.01494}, 
}

@inproceedings{
    swe-bench,
    title={{SWE}-bench: Can Language Models Resolve Real-world Github Issues?},
    author={Carlos E Jimenez and John Yang and Alexander Wettig and Shunyu Yao and Kexin Pei and Ofir Press and Karthik R Narasimhan},
    booktitle={The Twelfth International Conference on Learning Representations},
    year={2024},
    url={https://openreview.net/forum?id=VTF8yNQM66}
}

@misc{bug-category-DL,
      title={A Comprehensive Study on Deep Learning Bug Characteristics}, 
      author={Md Johirul Islam and Giang Nguyen and Rangeet Pan and Hridesh Rajan},
      year={2019},
      eprint={1906.01388},
      archivePrefix={arXiv},
      primaryClass={cs.SE},
      url={https://arxiv.org/abs/1906.01388}, 
}

@book{category-taxonomy-source,
  title={Software System Testing and Quality Assurance},
  author={Beizer, B.},
  isbn={9780442213060},
  lccn={83010458},
  series={Electrical-Computer Science and Engineering Series},
  url={https://books.google.com/books?id=zNAmAAAAMAAJ},
  year={1984},
  publisher={Van Nostrand Reinhold}
}

@standard{ISO25010,
  author       = {{International Organization for Standardization} and
                  {International Electrotechnical Commission}},
  title        = {Systems and software engineering --- Systems and software Quality Requirements and Evaluation (SQuaRE) --- Product quality model},
  number       = {ISO/IEC 25010:2023},
  year         = {2023},
  publisher    = {ISO},
  address      = {Geneva, Switzerland},
  url          = {https://www.iso.org/standard/78176.html}
}

@inproceedings{g-eval,
    title = "{G}-Eval: {NLG} Evaluation using Gpt-4 with Better Human Alignment",
    author = "Liu, Yang  and
      Iter, Dan  and
      Xu, Yichong  and
      Wang, Shuohang  and
      Xu, Ruochen  and
      Zhu, Chenguang",
    editor = "Bouamor, Houda  and
      Pino, Juan  and
      Bali, Kalika",
    booktitle = "Proceedings of the 2023 Conference on Empirical Methods in Natural Language Processing",
    month = dec,
    year = "2023",
    address = "Singapore",
    publisher = "Association for Computational Linguistics",
    url = "https://aclanthology.org/2023.emnlp-main.153/",
    doi = "10.18653/v1/2023.emnlp-main.153",
    pages = "2511--2522"
}

@misc{reflexion,
      title={Reflexion: Language Agents with Verbal Reinforcement Learning}, 
      author={Noah Shinn and Federico Cassano and Edward Berman and Ashwin Gopinath and Karthik Narasimhan and Shunyu Yao},
      year={2023},
      eprint={2303.11366},
      archivePrefix={arXiv},
      primaryClass={cs.AI},
      url={https://arxiv.org/abs/2303.11366}, 
}

@misc{gpt-5,
      title={OpenAI GPT-5 System Card}, 
      author={Aaditya Singh and Adam Fry and Adam Perelman and Adam Tart and Adi Ganesh and Ahmed El-Kishky and Aidan McLaughlin and Aiden Low and AJ Ostrow and Akhila Ananthram and Akshay Nathan and Alan Luo and Alec Helyar and Aleksander Madry and Aleksandr Efremov and Aleksandra Spyra and Alex Baker-Whitcomb and Alex Beutel and Alex Karpenko and Alex Makelov and Alex Neitz and Alex Wei and Alexandra Barr and Alexandre Kirchmeyer and Alexey Ivanov and Alexi Christakis and Alistair Gillespie and Allison Tam and Ally Bennett and Alvin Wan and Alyssa Huang and Amy McDonald Sandjideh and Amy Yang and Ananya Kumar and Andre Saraiva and Andrea Vallone and Andrei Gheorghe and Andres Garcia Garcia and Andrew Braunstein and Andrew Liu and Andrew Schmidt and Andrey Mereskin and Andrey Mishchenko and Andy Applebaum and Andy Rogerson and Ann Rajan and Annie Wei and Anoop Kotha and Anubha Srivastava and Anushree Agrawal and Arun Vijayvergiya and Ashley Tyra and Ashvin Nair and Avi Nayak and Ben Eggers and Bessie Ji and Beth Hoover and Bill Chen and Blair Chen and Boaz Barak and Borys Minaiev and Botao Hao and Bowen Baker and Brad Lightcap and Brandon McKinzie and Brandon Wang and Brendan Quinn and Brian Fioca and Brian Hsu and Brian Yang and Brian Yu and Brian Zhang and Brittany Brenner and Callie Riggins Zetino and Cameron Raymond and Camillo Lugaresi and Carolina Paz and Cary Hudson and Cedric Whitney and Chak Li and Charles Chen and Charlotte Cole and Chelsea Voss and Chen Ding and Chen Shen and Chengdu Huang and Chris Colby and Chris Hallacy and Chris Koch and Chris Lu and Christina Kaplan and Christina Kim and CJ Minott-Henriques and Cliff Frey and Cody Yu and Coley Czarnecki and Colin Reid and Colin Wei and Cory Decareaux and Cristina Scheau and Cyril Zhang and Cyrus Forbes and Da Tang and Dakota Goldberg and Dan Roberts and Dana Palmie and Daniel Kappler and Daniel Levine and Daniel Wright and Dave Leo and David Lin and David Robinson and Declan Grabb and Derek Chen and Derek Lim and Derek Salama and Dibya Bhattacharjee and Dimitris Tsipras and Dinghua Li and Dingli Yu and DJ Strouse and Drew Williams and Dylan Hunn and Ed Bayes and Edwin Arbus and Ekin Akyurek and Elaine Ya Le and Elana Widmann and Eli Yani and Elizabeth Proehl and Enis Sert and Enoch Cheung and Eri Schwartz and Eric Han and Eric Jiang and Eric Mitchell and Eric Sigler and Eric Wallace and Erik Ritter and Erin Kavanaugh and Evan Mays and Evgenii Nikishin and Fangyuan Li and Felipe Petroski Such and Filipe de Avila Belbute Peres and Filippo Raso and Florent Bekerman and Foivos Tsimpourlas and Fotis Chantzis and Francis Song and Francis Zhang and Gaby Raila and Garrett McGrath and Gary Briggs and Gary Yang and Giambattista Parascandolo and Gildas Chabot and Grace Kim and Grace Zhao and Gregory Valiant and Guillaume Leclerc and Hadi Salman and Hanson Wang and Hao Sheng and Haoming Jiang and Haoyu Wang and Haozhun Jin and Harshit Sikchi and Heather Schmidt and Henry Aspegren and Honglin Chen and Huida Qiu and Hunter Lightman and Ian Covert and Ian Kivlichan and Ian Silber and Ian Sohl and Ibrahim Hammoud and Ignasi Clavera and Ikai Lan and Ilge Akkaya and Ilya Kostrikov and Irina Kofman and Isak Etinger and Ishaan Singal and Jackie Hehir and Jacob Huh and Jacqueline Pan and Jake Wilczynski and Jakub Pachocki and James Lee and James Quinn and Jamie Kiros and Janvi Kalra and Jasmyn Samaroo and Jason Wang and Jason Wolfe and Jay Chen and Jay Wang and Jean Harb and Jeffrey Han and Jeffrey Wang and Jennifer Zhao and Jeremy Chen and Jerene Yang and Jerry Tworek and Jesse Chand and Jessica Landon and Jessica Liang and Ji Lin and Jiancheng Liu and Jianfeng Wang and Jie Tang and Jihan Yin and Joanne Jang and Joel Morris and Joey Flynn and Johannes Ferstad and Johannes Heidecke and John Fishbein and John Hallman and Jonah Grant and Jonathan Chien and Jonathan Gordon and Jongsoo Park and Jordan Liss and Jos Kraaijeveld and Joseph Guay and Joseph Mo and Josh Lawson and Josh McGrath and Joshua Vendrow and Joy Jiao and Julian Lee and Julie Steele and Julie Wang and Junhua Mao and Kai Chen and Kai Hayashi and Kai Xiao and Kamyar Salahi and Kan Wu and Karan Sekhri and Karan Sharma and Karan Singhal and Karen Li and Kenny Nguyen and Keren Gu-Lemberg and Kevin King and Kevin Liu and Kevin Stone and Kevin Yu and Kristen Ying and Kristian Georgiev and Kristie Lim and Kushal Tirumala and Kyle Miller and Lama Ahmad and Larry Lv and Laura Clare and Laurance Fauconnet and Lauren Itow and Lauren Yang and Laurentia Romaniuk and Leah Anise and Lee Byron and Leher Pathak and Leon Maksin and Leyan Lo and Leyton Ho and Li Jing and Liang Wu and Liang Xiong and Lien Mamitsuka and Lin Yang and Lindsay McCallum and Lindsey Held and Liz Bourgeois and Logan Engstrom and Lorenz Kuhn and Louis Feuvrier and Lu Zhang and Lucas Switzer and Lukas Kondraciuk and Lukasz Kaiser and Manas Joglekar and Mandeep Singh and Mandip Shah and Manuka Stratta and Marcus Williams and Mark Chen and Mark Sun and Marselus Cayton and Martin Li and Marvin Zhang and Marwan Aljubeh and Matt Nichols and Matthew Haines and Max Schwarzer and Mayank Gupta and Meghan Shah and Melody Huang and Meng Dong and Mengqing Wang and Mia Glaese and Micah Carroll and Michael Lampe and Michael Malek and Michael Sharman and Michael Zhang and Michele Wang and Michelle Pokrass and Mihai Florian and Mikhail Pavlov and Miles Wang and Ming Chen and Mingxuan Wang and Minnia Feng and Mo Bavarian and Molly Lin and Moose Abdool and Mostafa Rohaninejad and Nacho Soto and Natalie Staudacher and Natan LaFontaine and Nathan Marwell and Nelson Liu and Nick Preston and Nick Turley and Nicklas Ansman and Nicole Blades and Nikil Pancha and Nikita Mikhaylin and Niko Felix and Nikunj Handa and Nishant Rai and Nitish Keskar and Noam Brown and Ofir Nachum and Oleg Boiko and Oleg Murk and Olivia Watkins and Oona Gleeson and Pamela Mishkin and Patryk Lesiewicz and Paul Baltescu and Pavel Belov and Peter Zhokhov and Philip Pronin and Phillip Guo and Phoebe Thacker and Qi Liu and Qiming Yuan and Qinghua Liu and Rachel Dias and Rachel Puckett and Rahul Arora and Ravi Teja Mullapudi and Raz Gaon and Reah Miyara and Rennie Song and Rishabh Aggarwal and RJ Marsan and Robel Yemiru and Robert Xiong and Rohan Kshirsagar and Rohan Nuttall and Roman Tsiupa and Ronen Eldan and Rose Wang and Roshan James and Roy Ziv and Rui Shu and Ruslan Nigmatullin and Saachi Jain and Saam Talaie and Sam Altman and Sam Arnesen and Sam Toizer and Sam Toyer and Samuel Miserendino and Sandhini Agarwal and Sarah Yoo and Savannah Heon and Scott Ethersmith and Sean Grove and Sean Taylor and Sebastien Bubeck and Sever Banesiu and Shaokyi Amdo and Shengjia Zhao and Sherwin Wu and Shibani Santurkar and Shiyu Zhao and Shraman Ray Chaudhuri and Shreyas Krishnaswamy and Shuaiqi and Xia and Shuyang Cheng and Shyamal Anadkat and Simón Posada Fishman and Simon Tobin and Siyuan Fu and Somay Jain and Song Mei and Sonya Egoian and Spencer Kim and Spug Golden and SQ Mah and Steph Lin and Stephen Imm and Steve Sharpe and Steve Yadlowsky and Sulman Choudhry and Sungwon Eum and Suvansh Sanjeev and Tabarak Khan and Tal Stramer and Tao Wang and Tao Xin and Tarun Gogineni and Taya Christianson and Ted Sanders and Tejal Patwardhan and Thomas Degry and Thomas Shadwell and Tianfu Fu and Tianshi Gao and Timur Garipov and Tina Sriskandarajah and Toki Sherbakov and Tomer Kaftan and Tomo Hiratsuka and Tongzhou Wang and Tony Song and Tony Zhao and Troy Peterson and Val Kharitonov and Victoria Chernova and Vineet Kosaraju and Vishal Kuo and Vitchyr Pong and Vivek Verma and Vlad Petrov and Wanning Jiang and Weixing Zhang and Wenda Zhou and Wenlei Xie and Wenting Zhan and Wes McCabe and Will DePue and Will Ellsworth and Wulfie Bain and Wyatt Thompson and Xiangning Chen and Xiangyu Qi and Xin Xiang and Xinwei Shi and Yann Dubois and Yaodong Yu and Yara Khakbaz and Yifan Wu and Yilei Qian and Yin Tat Lee and Yinbo Chen and Yizhen Zhang and Yizhong Xiong and Yonglong Tian and Young Cha and Yu Bai and Yu Yang and Yuan Yuan and Yuanzhi Li and Yufeng Zhang and Yuguang Yang and Yujia Jin and Yun Jiang and Yunyun Wang and Yushi Wang and Yutian Liu and Zach Stubenvoll and Zehao Dou and Zheng Wu and Zhigang Wang},
      year={2025},
      eprint={2601.03267},
      archivePrefix={arXiv},
      primaryClass={cs.CL},
      url={https://arxiv.org/abs/2601.03267}, 
}

@misc{gpt-5.2,
  author       = {{OpenAI}},
  title        = {GPT-5.2},
  year         = {2025},
  howpublished = {\url{https://openai.com/index/introducing-gpt-5-2/}},
  note         = {Large language model}
}

@misc{claude-4.5-sonnet,
  author       = {{Anthropic}},
  title        = {Claude Sonnet 4.5},
  year         = {2025},
  howpublished = {\url{https://www.anthropic.com/news/claude-sonnet-4-5}},
  note         = {Large language model}
}

@inproceedings{bleu,
    title = "{B}leu: a Method for Automatic Evaluation of Machine Translation",
    author = "Papineni, Kishore  and
      Roukos, Salim  and
      Ward, Todd  and
      Zhu, Wei-Jing",
    editor = "Isabelle, Pierre  and
      Charniak, Eugene  and
      Lin, Dekang",
    booktitle = "Proceedings of the 40th Annual Meeting of the Association for Computational Linguistics",
    month = jul,
    year = "2002",
    address = "Philadelphia, Pennsylvania, USA",
    publisher = "Association for Computational Linguistics",
    url = "https://aclanthology.org/P02-1040/",
    doi = "10.3115/1073083.1073135",
    pages = "311--318"
}
